\documentclass[twocolumn]{aa}
\usepackage{graphicx}
\usepackage{txfonts}
\begin{document}
  \headnote{Letter to the Editor}
   \title{The ISO--2MASS AGN survey: On the type--1 sources
     \thanks{Based on
       observations with the Infrared Space Observatory
       ISO, an ESA project with instruments
       funded by ESA Member States (especially the PI countries: France,
       Germany, the Netherlands and UK) and
       with the participation of ISAS and NASA.
       The Two Micron All
       Sky Survey is a joint project of the University of
       Massachusetts and IPAC/Caltech, funded by the National
       Aeronautics and Space Administration and the National Science
       Foundation.
     }
   }

   \author{C. Leipski
          \inst{1}
          \and
          M. Haas
	  \inst{1}
	  \and
	  H. Meusinger
	  \inst{2}
          \and
	  R. Siebenmorgen
	  \inst{3}
          \and
	  R. Chini
	  \inst{1}
          \and
	  C. M. Scheyda
	  \inst{1}
	  \and\\
	  M. Albrecht
	  \inst{4}
	  \and
	  B. J. Wilkes
	  \inst{5}
          \and
	  J. P. Huchra
	  \inst{5}
          \and
	  S. Ott
	  \inst{6}
          \and
          C. Cesarsky
	  \inst{3}
          \and
	  R. Cutri
	  \inst{7}
	   }

   \offprints{Christian Leipski, leipski@astro.rub.de}

   \institute{Astronomisches Institut Ruhr--Universit\"at Bochum (AIRUB),
              Universit\"atsstra{\ss}e 150, 44780 Bochum, Germany
             \and
             Th\"uringer Landessternwarte Tautenburg (TLS), Sternwarte 5,
             07778 Tautenburg, Germany
             \and
             European Southern Observatory (ESO),
             Karl--Schwarzschild--Str. 2, 85748 Garching, Germany
             \and
	     Instituto de Astronom\'ia, Universidad Cat\'olica del
             Norte (UCN),  Avenida Angamos 0610, Antofagasta, Chile
             \and
             Harvard--Smithsonian Center for Astrophysics (CfA), 60 Garden Street, 
              Cambridge, MA 02138, USA
             \and
             HERSCHEL Science Centre, ESA, Noordwijk, PO Box 299, 2200
             AG Noordwijk, The Netherlands
	     \and
	     IPAC, California Institute of Technology
	     (Caltech),  770 South Wilson Avenue, Pasadena, CA
              91125, USA
   }
   \date{Received June 08, 2005; accepted July 09, 2005}

   \abstract{We combined the {\it ISOCAM Parallel Mode Survey} at
   6.7\,$\mu$m ($LW2$ filter) with the {\it Two Micron All
   Sky Survey} in order to obtain a powerful tool to search for AGN
   independent of dust extinction. 
   Using moderate colour criteria $H-K>0.5$ and $K-LW2>2.7$ we have
   selected a sample of 77 AGN candidates in an effective area of
   $\sim$10 square 
   degrees. By means of optical spectroscopy we find 24 ($\sim30\,\%$)
   type--1 QSOs at redshifts $0.1<z<2.3$; nine of them have $z>0.8$.
   About one third of the ISO--2MASS QSOs show so red 
   optical colours, that they are missed in optical and UV AGN surveys
   like SDSS, 2DF, or HES.
   With a surface density
   of about 2~deg$^{\rm - 2}$  down to $R<18$ mag
   the ISO--2MASS QSOs outnumber the 1.35~deg$^{\rm - 2}$ of the
   SDSS quasar survey by 50\%; we find a combined optical--IR QSO
   surface density of 2.7~deg$^{-2}$.
   Since only two of the ISO--2MASS QSOs have also $J-K>2$,
   the inclusion of the ISO mid--infrared photometry significantly extends
   the capabilities of the pure 2MASS red AGN survey. 
   We suggest that the newly found red AGN resemble
   young members of the quasar population, and that quasars spend much of their
   lifetime in a dust enshrouded phase.
   
   }
\titlerunning{The ISO--2MASS AGN}
\authorrunning{C. Leipski et\,al.}
   \maketitle
%

\section{Introduction}
Attempts to overcome the limits by dust extinction 
in optical AGN surveys and to identify
the entire AGN population -- including type--2 and buried AGN --
encompass surveys in the radio, X--ray and infrared (IR) ranges.
However, only about 30\% of
AGN are radio--loud (Urry \& Padovani \cite{urry95}). Hard X--rays enabled the
discoveries of elusive AGN completely hidden in starburst nuclei
(Maiolino et\,al. \cite{maiolino03}). However,
there exists also a significant fraction of 
X--ray faint AGN (Wilkes et\,al. \cite{wilkes02}),
suggesting that also other search techniques should be considered. 
The finding of obscured AGN is further complicated by
the contribution of the host galaxies, which may dominate the observed
properties.
Using IRAS 25\,$\mu$m/60\,$\mu$m colours far--IR searches already indicated
that the local space density of AGN may be significantly higher than
deduced from optical searches (Low et\,al. \cite{low88}).
Among far--IR dominant ULIRGs 
only few show AGN--typical mid--IR
spectral lines (e.g. Armus et\,al. \cite{armus04}) or 
X--ray evidence for powerful buried quasars (Ptak et\,al. \cite{ptak03}).
Searching among the 2MASS survey for very red AGN 
the extreme $J$-$K$$>$2 color selection reveals  new type--1 AGN
at redshifts $z < 0.8$ with moderate luminosities (Cutri
et\,al. \cite{cutri02}).  
The FIRST--2MASS study finds about 20\% previously overlooked
radio--loud quasars not suspicious in the UV (Glikman
et\,al. \cite{glikman04}). 
Although the contribution of the 2MASS red AGN to the
cosmic X-ray background may be as high as 30\%
(Wilkes et\,al. \cite{wilkes03}),
a considerable fraction of the AGN population might still be missed.\\\indent
The disadvantage of heavy extinction in optical surveys
can turn into a valuable detection tool, when observing
dust--surrounded AGN at near--infrared (NIR) {\it and} mid--infrared (MIR)
wavelengths. There, the reemission of
the dust heated by the strong radiation field of the AGN
should be seen as IR excess.
We  have started a new approach,
searching for AGN by means of their near-- {\it and} mid--IR
emission properties of the putative nuclear dust torus. 
The {\it ISOCAM Parallel Mode Survey} ``ISOCP" (Cesarsky
et\,al. \cite{cesarsky96}, 
Siebenmorgen et\,al. \cite{siebenmorgen96}, Ott et\,al. \cite{ott03},
Ott et\,al. \cite{ott05}) 
provides 6.7\,$\mu$m data for a
large number of extragalactic sources and is therefore an ideal
hunting ground for a hitherto unknown population of AGN.
The sample selection and first results from a 
subsample are described in detail by Haas et\,al. (\cite{haas04}).
Also other MIR searches have been started using the Spitzer Space
Telescope (e.g. Lacy et\,al. \cite{lacy04}).
Here we report on the results for type--1 AGN from the full sample
of those ISOCP sources which have 2MASS counterparts.
\begin{figure}
  \resizebox{\hsize}{!}{\includegraphics[angle=0]{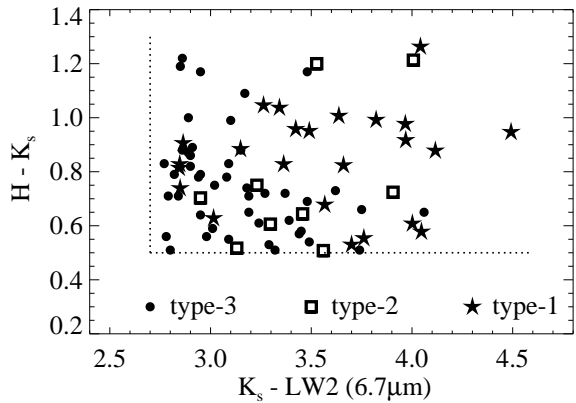}}
  \caption{IR colour--colour diagram of the ISO--2MASS AGN.  
  \label{types_in_cc}
  }
\end{figure}
\section{Data}
From a sample of 3000 high galactic latitude 
($|b|$$>$20$^{\circ}$) sources detected on
randomly distributed frames covering a total effective area of
$\sim$10\,deg$^{2}$ we have found
unresolved (FWHM\,$\sim$$6\arcsec$) objects with steep
2.2-6.7\,$\mu$m slopes, which we consider as AGN candidates. 
By means of correlations with the 2MASS archive
and by comparison with colour--colour and colour--magnitude properties of
known sources we have excluded -- as far as possible --
contaminations like stars or pure star forming galaxies
(Haas et\,al. \cite{haas04}).\\\indent
The selection criterion for the ISO--2MASS AGN is a good detection
in the ISO LW2 filter 
down to F$_{\rm 6.7 \mu m}$ $\sim$ 1\,mJy as well as in all 2MASS filters, $J,
H,$ and $K_s$, respectively. In addition to these flux limits we apply,
guided by the comparison with PG quasars and 3CR radio galaxies, 
only the moderate colour criteria $H-K_s>0.5$ {\sl and} $K_s-{\rm LW2}>2.7$
(Vega--based system). 
By this procedure 77 candidates were selected,  of which eight had
redshifts available in the NED. For the remaining 69 sources we have
performed optical spectroscopy at various telescopes.
\section{Results and discussion}
\subsection{Properties of the ISO--2MASS AGN}
\begin{figure}
  \resizebox{\hsize}{!}{\includegraphics[angle=0]{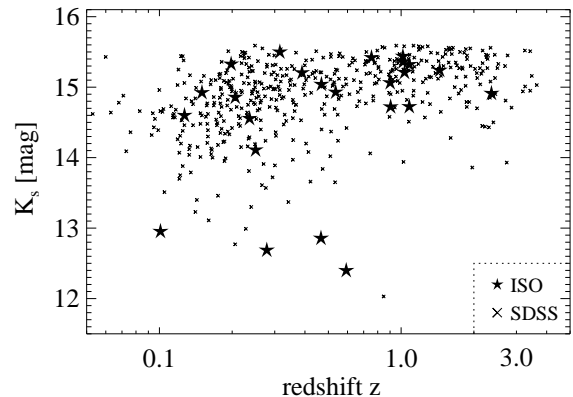}}
  \caption[]{\label{k_over_z}Distribution of $K_s$ magnitude
    and redshift
    for the ISO--2MASS type--1 AGNs and SDSS--QSOs (only every
    tenth object plotted).}
\end{figure}
\begin{figure}
  \vspace{-4mm}
  \resizebox{\hsize}{!}{\includegraphics{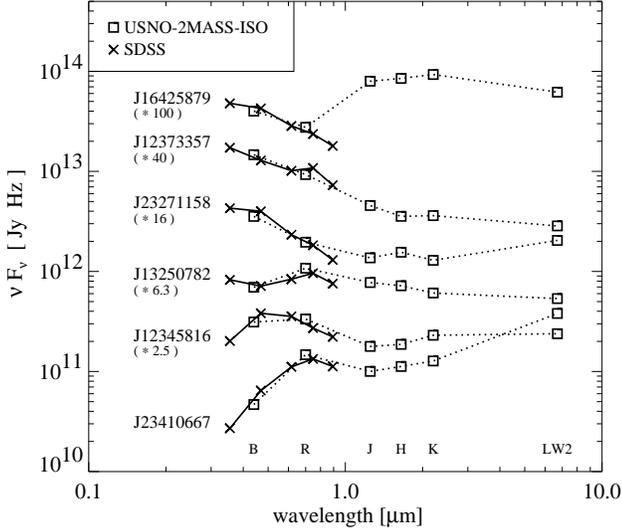}}
  \caption[]{\label{seds}SEDs of those six ISO--2MASS QSOs with SDSS
  photometry. Dotted lines refer to USNO--B photometry and solid lines
  to SDSS photometry. The good match confirms the USNO data.}
\end{figure}
\begin{figure}
  \vspace{-4mm}
  \resizebox{\hsize}{!}{\includegraphics{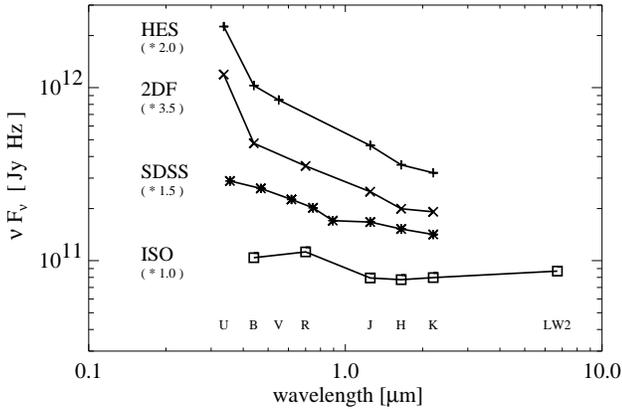}}
  \caption[]{\label{mean_sed} Mean SEDs of QSO samples with 2MASS
  counterparts.} 
\end{figure}

Within our sample we find 24 broad-line type--1 AGN 
($\sim$31\,\%, redshift range z=0.1--2.3), nine narrow-line
type--2 AGN ($\sim$12\,\%,
z=0.1--0.3), and 44 emission line galaxies with LINER and HII
type spectra ($\sim$57\,\%, z=0.03--0.3).
None of the objects turned out to be a star. The emission line
galaxies, henceforth denoted type--3 sources, 
are heavily reddened (H$_{\alpha}$/H$_{\beta}$\,$>$\,10) and their
spectra show clear signatures of the host galaxy. Their
high MIR/NIR, but low FIR/MIR flux ratio typical for AGN argues
against pure starbursts. Essentially none of the
sources has been detected by IRAS.
The distribution of the different types of sources in the
colour--colour diagram is shown in Fig.\,\ref{types_in_cc}.
While in $H-K_s$ only minor trends are present, we see a striking
dependence in $K_s-LW2$: 
The type--1 and type--3
sources concentrate toward the right-- and left--hand sides,
respectively, while
type--2 sources are more intermediate.
This suggests that we see the hot dust emission best in the type--1 sources,
while it is more obscured or intrinsically less prominent in some of the
type--2 sources and in most of the type--3 ones.
The type--2 and type--3 sources will be investigated in detail in
  a forthcoming paper.
In the following discussion we consider only the type--1 sources and
Tab.\,\ref{type1_sources} summarises their parameters.\\\indent 
The $K_s$ brightness of the type--1 AGN spans the range $12.4 < K_s < 15.5$.
Figure\,\ref{k_over_z} shows the distribution of $K_s$ over $z$. 
Using a $\Lambda$ cosmology with 
H$_0$ = 71 km\,s$^{-1}$\, Mpc$^{-1}$, $\Omega_{{\rm matter}}$ = 0.27
and $\Omega_{\Lambda}$ = 0.73,
the type--1 sources exhibit an absolute $K_s$-band magnitude in the range
of $-25$ to $-30$, similar to the SDSS quasars. This qualifies them as QSOs,
henceforth denoted ISO--2MASS QSOs. In this calculation no
$k$--correction was applied; if done, it would further increase the
luminosity of the objects.
Five of the ISO--2MASS QSOs are detected by NVSS or FIRST, three being
radio--loud with F$_{\rm 1.4 GHz}$ $>$ F$_{\rm 2.2 \mu m}$.\\\indent
Optical $B$-- and $R$--band photometry of the sources is provided by the USNO
catalogue (USNO--B, Monet et\, al. \cite{monet03}),
with a range of $15.7 < B < 19.7 $ and $15.5<R<17.9$.
We found that the $B$ and $R$--band photometry is consistent with that
derived from the spectra. 
The ISO--2MASS AGN span a colour range $-0.4<B-R<2.2$;
$42$\,\%  (10/24) have $B-R>1$.
Figure\,\ref{seds} shows the MIR to optical 
spectral energy distributions (SEDs) for those sources for which also
SDSS photometry is available. Even the mean SED shows red colours
 compared to other samples, especially at shorter wavelengths
 (Fig.\,\ref{mean_sed}).   
\subsection{Comparison with optical-UV selected QSOs}
\begin{figure}
  \resizebox{\hsize}{!}{\includegraphics{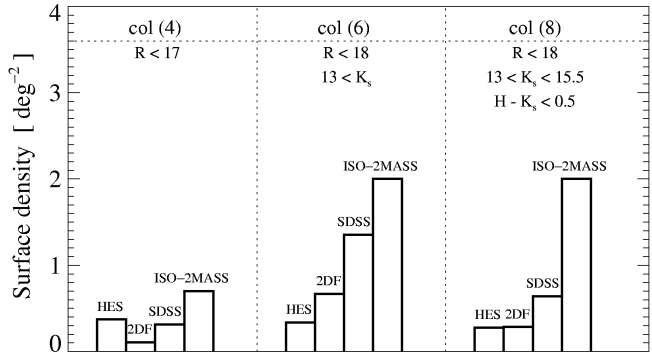}}
  \caption[]{\label{number_histo} QSO number counts of the ISO--2MASS
  and three optical surveys at three different bins. The column
  enumeration refers to Tab.\,\ref{number_counts}.}
\end{figure}

If the ISO--2MASS QSOs comprise the same QSO population as that found
by optical-UV selected QSO samples, then the number counts
as well as the mean SEDs should be similar for suitably matched
bins. 
We compare the ISO--2MASS QSOs with the quasars in
the SDSS DR3 (Schneider et\,al. \cite{schneider05}),
the 2QZ$+$6QZ catalogues
of the 2DF survey (Croom et\,al. \cite{croom04}), and the
Hamburg/ESO quasar survey (HES, Wisotzki
et\,al. \cite{wisotzki00}). We also correlated these reference
catalogues with the 2MASS archive, thereby creating sub--samples
hereafter called SDSS--2MASS, 2DF--2MASS, and HES--2MASS, respectively.
Fig.\,\ref{k_over_z} shows the redshift
and $K_s$--band magnitude distributions of the ISO--2MASS QSOs and the
SDSS--2MASS QSOs. 
Apart from the low number statistics of the ISO--2MASS QSOs, we find that
the redshift and $K_s$--band magnitude
distributions of the ISO--2MASS QSOs and all three optical
samples do not differ severely, so that a comparison of the number
counts makes sense. However, the SDSS spectroscopy is limited to
  sources with $i>15$ (Richards et\,al. \cite{richards02}) which
  translates to $K_s>13$ (only $\sim$10\%~of the SDSS--2MASS QSOs have
$i-K_s<2$). Therefore we exclude all objects with $K_s<13$ in the
  following discussion.\\\indent
In order to compare the number of quasars found per deg$^2$ for the
  different samples we chose
various bins down to the flux limits of the 
ISO--2MASS QSOs at $R<18$ and $K_s<15.5$ and separate also at $z = 0.8$. 
The USNO photometry yields on average smaller $B-R$ values
compared with newer photometric samples, mainly
because of differences in the $B$--band, while the $R$-band values are more
comparable. 
In order to allow for a more homogeneous photometric comparison,
we also used the $R$--band photometry from USNO
for selecting the optical QSO sub--samples
(Table\,\ref{number_counts}). The basic results are illustrated in
Fig.\,\ref{number_histo}. The striking result is that for all resonable bins 
the surface density of ISO--2MASS QSOs is
by a factor of 1.5 to 10 higher than for the optically
selected QSOs. We did not find any reasonable bins to match the
surface densities of the IR-- and the optically
selected QSOs samples.\\\indent
\begin{table}
\caption{Parameters of the ISO--2MASS type--1 QSOs.}             
\label{type1_sources}      
\centering                          
{\footnotesize
\begin{tabular}{lrrl}        
\hline\hline                 
2MASS               & F$_{\rm 6.7 \mu m}$ &redshift& remarks \\    
                    & [mJy] &        &         \\    
\hline                                        
&& \multicolumn{2}{l}{~~~~$z$ \& type from this work}   \\
J01363451+4112497   &  1.21 & 0.198  & \\
J03005029$-$7938450 &  3.53 & 0.901  & \\
J03080561$-$6535520 &  1.79 & 0.127  & \\
J11323474$-$1952448 &  1.65 & 0.389  & \\
J11353051$-$1425344 &  1.94 & 0.536  & $J$$-$$K_s$$>$$2$ \\
J13250782+0541052   &  1.65 & 0.206  & SDSS phot+spec\\
J13474549$-$0841059 &  1.43 & 0.754  & \\
J14265292+3323231   &  2.59 & 1.083  & \\
J16413748+6541140   &  1.60 & 1.082  & \\
J17075650+5630136   &  1.75 & 1.019  & \\
J17175947+4956261   &  2.07 & 1.018  & \\
J20145083$-$2710429 &  2.44 & 1.444  & \\
J22484115$-$5109532 & 16.90 & 0.101  & \\
J23271158+0114469   &  3.46 & 0.468  & SDSS phot\\
J23324586$-$1414242 &  2.51 & 0.315  & \\
J23410667$-$0914327 &  8.49 & 0.236  & SDSS phot\\
\hline		    	     	     		    
&& \multicolumn{2}{l}{~~~~$z$ \& type from NED}   \\
J00300421$-$2842259 & 31.35 & 0.278  & RQ, $J$$-$$K_s$$>$$2$ \\
J01020053$-$3018259 &  2.24 & 1.033  & RL \\
J12345816+1308549   &  2.12 & 2.364  & SDSS phot\\
J12373357+1319063   &  1.33 & 0.150  & SDSS phot\\
J14590760+7140199   &  3.11 & 0.905  & RL (3C\,309.1)\\
J15520240+2014016   &  5.09 & 0.250  & \\
J16425879+3948369   & 13.82 & 0.593  & RL (3C\,345) \\
                    &       &        & SDSS phot+spec\\
J21145258+0607423   & 25.04 & 0.466  & RQ \\
\hline                                   
\end{tabular}
}
\end{table}
This result is remarkable as Vanden Berk
et\,al. (\cite{vandenberk05}) report a completeness of 80\%~to
95\% for the SDSS quasar survey.
We searched the SDSS DR3 for photometric and
spectroscopic data of the 24 ISO--2MASS QSOs. Six have
$ugriz$ photometry available; compared to the mean SEDs
  (Fig.\,\ref{mean_sed}) they show a more or
less red SED, even shortward of the B--band (Fig.\,\ref{seds}).  
According to the SDSS colour criteria
(Richards et\,al. \cite{richards02}) two of these six sources lie in the
stellar loci and are not foreseen for SDSS spectroscopy,
two seem to be potential QSO candidates
and two have
been identified spectroscopically  as QSO.
The extrapolation from these six sources indicates
that the completion of the SDSS spectroscopy may at most double the
optical colour 
selected number of QSOs, and that one third of the 24 ISO--2MASS QSOs
will be missed by the SDSS spectroscopic QSO search
due to star--like colours. \\\indent
Figure\,\ref{mean_sed} illustrates that shortward of the
$R$-band the mean SED of the ISO--2MASS QSOs is significantly redder than
that of the optically selected QSOs (with 2MASS counterparts), in
particular for the 2DF--2MASS and the 
HES--2MASS QSOs, which show a strong upturn shortward of the B-band.
Both results, the
higher QSO surface density and the redder SEDs, are
independent of the magnitude or NIR colour bins chosen. We
conclude that 
the ISO--2MASS AGN survey discovers a QSO population, about a third of which
is clearly different from that found in the optical surveys.\\\indent
On the other hand, down to $R<18$ the 2DF and SDSS QSO surveys find
about 40-50\% blue QSOs which have $K_s > 15.5$, hence are fainter than
the detection limit of the ISO--2MASS survey; 
these optical QSOs without 2MASS counterpart have on average bluer
optical colours 
than those with 2MASS counterparts.\footnote{
  Notably, the ISOCAM Parallel Survey reveals 
  numerous ($>$100) $R>18$ sources without 2MASS counterpart, which
  are not considered here.
  Their NIR and optical investigation is still ongoing.} 
To get an estimate of the entire IR-- and optical QSO number
  counts down to 
$R<18$ we add the surface density of ISO--2MASS and SDSS QSOs and
subtract the intersection of both samples, i.e. those
SDSS quasars that also fulfill our IR selection criteria ($R<18$ \&
$13<K_s<15.5$ \& $H-K_s>0.5$).
Referring to columns 6 and 8 of Tab.\,\ref{number_counts} this corresponds to 
$\frac{20}{10}+\frac{5669}{4188}-\frac{2685}{4188}\approx2.7$~deg$^{-2}$
($\sim$\,1.5 deg$^{\rm -2}$ for $z>0.8$, respectively),
i.e. about a factor 2 higher than inferred from the SDSS
QSO survey alone.\\\indent
The fact that IR counts essentially add to the quasar surface density can
most likely be ascribed to 
quasars (extended as well as pointlike objects) that have stellar
colours. Remarkably, in the completeness test of the SDSS QSO survey
by Vanden Berk et\,al. (\cite{vandenberk05}) this population of
quasars has largely been excluded. However our data show that these
quasars with optical stellar--like colours comprise a
considerable fraction of the total population of quasars and that they can
most efficiently be discovered by IR colours.  
\subsection{Comparison with the 2MASS red AGN survey}
Using the colour selection $J-K_s>2$
the 2MASS red AGN survey found an extrapolated surface density of
$\sim$0.57 type--1
and type--2 AGN per deg$^{2}$ (Cutri et\,al. \cite{cutri02}), which
become lower based on newer larger data sets (Cutri, priv. com.).
 Two type--1  and two type--2 ISO--2MASS AGN match the
criterion $J-K_s>2$, resulting in 4 / 10 = 0.40 AGN per deg$^{2}$, 
roughly comparable to the 2MASS red AGN estimates.
Thus the 2MASS red AGN are a proper subset of the
ISO--2MASS AGN survey, as expected.\\\indent
Due to $k$--correction effects
the 2MASS red AGN survey is biased against sources with redshifts $z>0.8$,
hence against high luminosity sources (Cutri et\, al. \cite{cutri02}).
 Using a
moderate colour criterion $H-K_s > 0.5$ (roughly
corresponding to $J-K_s > 1.2$ as used by Francis
et\,al. \cite{francis04}) the ISO--2MASS AGN survey 
in fact finds nine (out of 24) QSOs with $z>0.8$.
As a consequence the ISO--2MASS--QSOs reach by one to three magnitudes
higher $K_s$ band luminosities.
\begin{table*}[t!]
\caption{Number counts of the IR and optical QSO samples for various bins.
  We adopt Poisson errors
  ($\sqrt{\rm N}$) for the ISO--2MASS QSO sample.
  In the upper four blocks of the table we use the $R$-band photometry from the
  USNO catalog, in the lower two ones the $R$-band photometry is from
  2DF and SDSS themselves (for SDSS: $R = g' - 1.14 (g' - r') - 0.14$
  according to Smith et\, al. \cite{smith02}).  
}
\label{number_counts}
\centering
{\footnotesize
\renewcommand{\footnoterule}{}  
\begin{tabular}{l|c|r|rr|rr|rr|rr|rr}
\hline \hline
\multicolumn{1}{c|}{(1)} & (2) & \multicolumn{1}{c}{(3)} & \multicolumn{2}{c}{(4)} & \multicolumn{2}{c}{(5)} & \multicolumn{2}{c}{(6)} & \multicolumn{2}{c}{(7)} & \multicolumn{2}{c}{(8)}\\ 
\multicolumn{1}{c|}{Sample} &  area & \multicolumn{11}{c}{number (N) and surface
density N/area (deg$^{-2}$) of quasars with} \\
 &  & total & \multicolumn{2}{c|}{$R<17$} & \multicolumn{2}{c|}{$R<18$} & \multicolumn{2}{c|}{$R<18$} & \multicolumn{2}{c|}{$R<18$} & \multicolumn{2}{c}{$R<18$} \\
 & & & \multicolumn{2}{c|}{} & \multicolumn{2}{c|}{} & \multicolumn{2}{c|}{$K_s>13$} & \multicolumn{2}{c|}{$13<K_s<15.5$} & \multicolumn{2}{c}{$13<K_s<15.5$}\\
& & & \multicolumn{2}{c|}{} & \multicolumn{2}{c|}{} & \multicolumn{2}{c|}{} & \multicolumn{2}{c|}{} & \multicolumn{2}{c}{$H-K_s>0.5$}\\
\hline
 & deg$^{2}$ & \multicolumn{1}{c|}{N} & \multicolumn{1}{c}{N} & \multicolumn{1}{c|}{deg$^{-2}$} & \multicolumn{1}{c}{N} & \multicolumn{1}{c|}{deg$^{-2}$} & \multicolumn{1}{c}{N} & \multicolumn{1}{c|}{deg$^{-2}$} & \multicolumn{1}{c}{N} & \multicolumn{1}{c|}{deg$^{-2}$} & \multicolumn{1}{c}{N} & \multicolumn{1}{c}{deg$^{-2}$} \\
\hline
\multicolumn{2}{c}{} & \multicolumn{11}{c}{$R$--band photometry from USNO}\\
\hline
ISO--2MASS    	      &           &       &      &         &      &         &      &               &      &         &      &               \\ 
$z>0.0$               & $\sim$10  &    24 &    7 &     0.7 &   24 &     2.4 &   20 &     {\bf 2.0} &   20 &     2.0 &   20 &     {\bf 2.0} \\
$z>0.8$	              & $\sim$10  &     9 &    1 &     0.1 &    9 &     0.9 &    9 &     {\bf 0.9} &    9 &     0.9 &    9 &     {\bf 0.9} \\
\hline		      		   	    	    				      
HES  		      &           &       &      &         &      &         &      &               &      &         &      &               \\
$B<17$                &   $>$1000 &   415 &  371 & $<$0.37 &  415 & $<$0.42 &  337 & {\bf $<$0.34} &  336 & $<$0.34 & 277 & {\bf $<$0.28} \\
$B<17$ and  $z>0.8$   &   $>$1000 &   140 &  122 & $<$0.12 &  140 & $<$0.14 &  137 & {\bf $<$0.14} &  137 & $<$0.14 &  84 & {\bf $<$0.08} \\
\hline		      		   	    	    		
SDSS DR3	      &           &       &      &         &      &         &      &               &      &         &      &               \\
$z>0.0$               & 4188      & 44298 & 1309 &    0.31 & 5713 &    1.36 & 5669 &    {\bf 1.35} & 3149 &    0.75 & 2685 &    {\bf 0.64} \\
$z>0.8$               & 4188      & 35459 &  497 &    0.12 & 3303 &    0.79 & 3303 &    {\bf 0.79} & 1213 &    0.29 &  846 &    {\bf 0.20} \\
\hline		      		   	    	    		
2DF 2QZ+6QZ	      &           &       &      &         &      &         &      &               &      &         &      &               \\
$z>0.0$               & 721.6     & 19304 &   76 &    0.11 &  484 &    0.67 &  483 &    {\bf 0.67} &  253 &    0.35 &	 205 &  {\bf 0.28} \\
$z>0.8$		      & 721.6     & 16055 &   41 &    0.06 &  320 &    0.44 &  320 &    {\bf 0.44} &  143 &    0.20 &   92 &    {\bf 0.13} \\
\hline		      		 	           			  			       
\multicolumn{2}{c}{} &  \multicolumn{11}{c}{$R$-band photometry from SDSS and 2DF, respectively}\\
\hline		      		 	           			  			       
SDSS DR3	      &           &       &      &         &      &         &      &               &      &         &      &               \\
$z>0.0$               & 4188      & 46420 &  568 &    0.14 & 4700 &    1.13 & 4657 &    {\bf 1.11} & 2653 &    0.63 & 2212 &    {\bf 0.53} \\
$z>0.8$               & 4188      & 37322 &  300 &    0.07 & 3068 &    0.73 & 3068 &    {\bf 0.73} & 1285 &    0.31 &  901 &    {\bf 0.22} \\
\hline		      		    			     
2DF 2QZ+6QZ	      &           &       &      &         &      &         &      &               &      &         &      &               \\
$z>0.0$               & 721.6     & 23660 &   62 &    0.09 &  727 &    1.01 &  726 &    {\bf 1.01} &  327 &    0.45 &  271 &    {\bf 0.38} \\
$z>0.8$		      & 721.6     & 19775 &   32 &    0.04 &  486 &    0.67 &  486 &    {\bf 0.67} &  161 &    0.22 &  115 &    {\bf 0.46} \\
\hline
\end{tabular}
}
\end{table*}
\section{The nature of the ISO-2MASS type--1 QSOs}
Combining the
ISO 6.7\,$\mu$m and 2MASS surveys we applied a
moderate near-- and mid--IR 
colour criterion to search for AGN.
About 30\% of the selected sources turned out to be type--1 QSOs. Part
of them have colour properties 
similar to optically selected QSOs, but about 30\% of them have red optical
SEDs similar to stars, so that they might escape QSO identification in
current optical colour surveys.\\\indent
In the frame work of a quasar's evolution from an initially
dust--enshrouded object to a clean one (Sanders et\,al. \cite{sanders88}, Haas
et\,al. \cite{haas03}) we suggest that the red objects comprise young members 
of the QSOs population. If true,
then the high (about 30\%) fraction of these young objects
indicates that the QSOs spend much of their life time in a dust
surrounded phase, before they change their appearance becoming
optically blue. Future studies may provide further clues to this issue
as well as their contribution to the X--ray background.
\begin{acknowledgements}
  Observing time for spectroscopy has been granted at the telescopes:
  Tautenburg 2--m, SAAO 1.9--m, CTIO 4--m, KPNO 2.1--m,
  ESO  NTT  3.5--m, CAHA 2.2--m, NOT 2.5--m and the TNG 3.5--m.
  Part of this work was supported by  
  \emph{Deut\-sche For\-schungs\-ge\-mein\-schaft, DFG\/} project
  number Ts~17/2--1, and by 
  Nordrhein--Westf\"alische Akademie der Wissenschaften. We
  thank the referee Roberto Maiolino for his critical expertise.
\end{acknowledgements}


\begin{thebibliography}{}
\bibitem[2004]{armus04} Armus, L., Charmandaris, V., Spoon, H.,
  et\,al. 2004, \apjs, 154, 178
\bibitem[1996]{cesarsky96} Cesarsky, C. J., Abergel, A., Agnese, P.,
  et\,al. 1996, \aap, 315, L32
\bibitem[2004]{croom04} Croom, S. 
    , Smith, R. 
    , Boyle, B. 
    , et\,al. 
    2004, \mnras, 349, 1397 
\bibitem[2002]{cutri02} Cutri, R. 
    , Nelson, B. 
    , Francis, P.
    , Smith, P. 
  2002, ASP 284, 127 
\bibitem[2004]{francis04} Francis, P., Nelson, B., Cutri, R. 2004,
  \aj, 127, 646 
\bibitem[2004]{glikman04} Glikman, E., Gregg, M. D., Lacy, M., et
  al. 2004, \apj, 607, 60
\bibitem[2003]{haas03} Haas, M., Klaas, U., M\"uller, S. A. H.,
  et\,al. 2003, \aap, 402, 87  
\bibitem[2004]{haas04} Haas, M., Siebenmorgen, R., Leipski, C., et
  al. 2004, \aap, 419, L49 
\bibitem[2004]{lacy04} 	Lacy, M., Storrie--Lombardi, L., Sajina,
  A., et\,al. 2004, \apjs, 154, 166
\bibitem[1988]{low88} Low, F., Cutri, R., Huchra, J., \& Kleinmann,
  S. 1988, \apj, 327, L41
\bibitem[2003]{maiolino03} Maiolino, R., Comastri, A., Gilli,
  R., et\,al. 2003,  MNRAS 344, L59 
\bibitem[2003]{monet03} Monet, D. G., Levine, S. E., Canzian, B., et
  al. 2003, \aj, 125, 984
\bibitem[2003]{ott03} Ott, S., Siebenmorgen, R., Schartel, N.,
  et\,al. 
  2003, ESA SP-511, 159
\bibitem[2005]{ott05} Ott, S., Siebenmorgen, R., Schartel, N., et
  al. 2005, \aap, submitted
\bibitem[2003]{ptak03}  Ptak A., Heckman T., Levenson
  N. A., et\,al. 2003, \apj 592, 782
\bibitem[2002]{richards02} Richards, G. T., Fan, X., Newberg, H. J.,
  et\,al. 2002, \aj, 123, 2945
\bibitem[1988]{sanders88}  Sanders, D., Soifer, T., Elias,
  J., et\,al. 1988, \apj 325, 74
\bibitem[2005]{schneider05} Schneider, D., Hall, P., Richards, G.,
  et\,al. 2005, astro--ph/0503679 
\bibitem[1996]{siebenmorgen96} Siebenmorgen, R., Abergel, A.,
  et\,al. 1996, \aap, 315, L169
\bibitem[2002]{smith02} Smith J. A., Tucker D., Kent S., et\,al. 2002,
  AJ 123, 2121 
\bibitem[1995]{urry95} Urry, C. M., \& Padovani, P. 1995, \pasp, 107, 803  
\bibitem[2005]{vandenberk05} Vanden\,Berk, D., Schneider D., Richards
  G., et\,al. 2005, \aj, 129, 2047
\bibitem[2002]{wilkes02} Wilkes, B. J., Schmidt, G. D., et\,al. 
   2002, \apj, 564, L65
\bibitem[2003]{wilkes03} Wilkes, B. J., Risaliti, G., Ghosh, H et\,al.
  2003, BAAS 203, 6304
\bibitem[2000]{wisotzki00} Wisotzki, L., Christlieb, N., Bade, N., et\,al. %
  2000, \aap, 358, 77

 
\end{thebibliography}
\end{document}